\documentclass{article}

\usepackage{microtype}
\usepackage{graphicx}
\usepackage{subfigure}
\usepackage{booktabs} % for professional tables
\usepackage{amsfonts}
\usepackage{amssymb}
\usepackage{graphicx}
\usepackage{amsmath}
\usepackage{endnotes}
\usepackage{setspace}
\usepackage{fancyvrb}
\usepackage{tikz}
\usetikzlibrary{arrows}

\usepackage{color}
\newcommand{\Qidone}[1]{\textcolor[rgb]{0.00,0.00,0.00}{#1}}

\usepackage{hyperref}

\usepackage[accepted]{icml2019}

\icmltitlerunning{Self Organizing Supply Chains for Micro-prediction}

\begin{document}

\twocolumn[
\icmltitle{Self Organizing Supply Chains for Micro-Prediction: \\
  Present and Future uses of the ROAR Protocol  }



\icmlsetsymbol{equal}{*}

\begin{icmlauthorlist}
\icmlauthor{Peter Cotton}{jpm}
\end{icmlauthorlist}

\icmlaffiliation{jpm}{JP Morgan Chase \& Co., New York, NY, United States}
\icmlcorrespondingauthor{Peter Cotton}{peter.d.cotton@jpmorgan.com}

\icmlkeywords{Machine Learning, ICML}

\vskip 0.3in
]



\printAffiliationsAndNotice{}  % leave blank if no need to mention equal contribution

\begin{abstract}
A multi-agent system is trialed as a means of crowd-sourcing inexpensive but high quality streams of predictions. Each agent is a micro-service embodying statistical models and endowed with economic self-interest. The ability to fork and modify simple agents is granted to a large number of employees in a firm and empirical lessons are reported. We suggest that one plausible trajectory for this project is the creation of a ``Prediction Web''. 
\end{abstract} 

\section{Motivating a Prediction Web} 

Under a plethora of labels, Applied Statistics has taken on new significance in the 21st Century. It might well be said that we live in a ``micro-prediction'' economy as our myriad movements, such as our locations and decisions in physical or commercial worlds, are either ostensibly predicted or indirectly forecast inside engines for recommendation, identification, pricing or navigation.

Micro-predictions, which we define as streams of repeated predictions of the same type in great number, drive real-time decision making and with it modern commerce - not to mention scientific and civic applications.\footnote{By using the phrase micro-prediction we distance the notion of thousands or millions of repeated predictions {\em of the same type} from the usual lay interpretation of prediction - referring to such things as next year's GDP or the outcome of an election.} A stream of micro-predictions is an economic good. Granted, this good is almost invariably coupled to a specific application or tool - but setting this aside the existence of any society in which a good is manufactured and distributed begs the central economic question: how to organize the production of prediction and the dissemination of goods to all members of society in an efficient manner?  

The present undertaking is premised on the idea that unbundling and commodification of prediction streams has the potential to unlock the efficacy of price propagation. We propose a micro-prediction economy centred around a simple protocol to be described. In our economy trade can be fully automated and economic agents take the form of intelligent microservices (``bots'' hereafter). These bots are authored by engineers and data scientists. 

\section{Background: Crowd-sourcing and the Common Task Framework}
\label{submission}

While operationally different to other data science crowd-sourcing efforts, ours is motivated by the success of prior work. The use of well defined quantitative tasks to identify talent and foster research is related and well appreciated. The important role of contests is summarized in \cite{Donoho201750Science} where the terminology Common Task Framework (CTF) from Marc Liberman is used to describe the following setup:

\begin{enumerate}
\item A publicly available training dataset involving, for each observation, a list of (possibly many)
feature measurements, and a class label for that observation.
\item A set of enrolled competitors whose common task is to infer a class prediction rule from the
training data.
\item A scoring referee, to which competitors can submit their prediction rule. The referee runs the
prediction rule against a testing dataset which is sequestered behind a Chinese wall. The
referee objectively and automatically reports the score (prediction accuracy) achieved by the
submitted rule.
\end{enumerate}

Academic data science contests and standardized data sets have a long history. There have been notable recent attempts to industrialize the paradigm. Kaggle.Com plays the role of scoring referee precisely as Liberman described, and with success \cite{Carpenter2011MayWin}. There, AllState's use of crowd-sourcing to improve their actuarial models is a cited win. The competition attracted six hundred data scientists and three hundred teams vying for \$10,000 in prizemoney. The contestants out-predicted Allstate's internal experts drawn from their actuarial department. 

General Electric reported similar gains when, in 2012, they offered a \$500,000 prize for prediction of airline flight arrival times. Close to 4,000 separate algorithms were entered and the crowd obliterated the industry benchmarks, reducing error by over $40$ percent.   

TopCoder is another site where algorithmic crowd-sourcing is facilitated. Harvard Medical School enlisted TopCoder to improve an edit distance calculation for DNA strings. 
\begin{quotation}
Prior to the TopCoder contest, the best known solution, MegaBLAST, processed 100,000 sequences to a high degree of accuracy, yet required $2,000$ seconds to execute. A Harvard provided benchmark that spent a year on this unique problem was able to produce an improved outcome, reducing the computational time to $400$ seconds.\footnote{https://www.topcoder.com/case-studies/harvard/} 
\end{quotation}
The contest reduced computation time to 16 seconds.  

An incomplete list of statistical crowd-sourcing communities is provided in Table \ref{tab:contests} and described further in \cite{GarciaMartinez2014TheAnalysis}. These efforts, and others not listed, have identified a rapidly growing pool of talent empowered by open source tooling and open education. 

\begin{table}[h!]
\label{tab:contests}
\begin{tabular}{ |l|l|l| }
\hline 
  Platform & Registered Users & Focus \\
  \hline
  TopCoder & $1,200,000$\footnote{https://en.wikipedia.org/wiki/Topcoder} & Programming \\
  Kaggle & $600,000$\footnote{https://www.kaggle.com/host/business}  & Data science \\
  Quantopian & $210,000$\footnote{https://en.wikipedia.org/wiki/Quantopian} & Quant trading \\
  QuantConnect & $75,000$\footnote{https://www.quantconnect.com/forum/discussions/1/interesting} & Quant trading \\
  CrowdAnalytix & $20,000$\footnote{https://www.crowdanalytix.com/community} & Data science \\
  \hline
\end{tabular}
\caption{Quantitative contest communities of approximately known size. We were not able to find estimates for WorldQuant, Numerai, Quantiacs and apologize for any omissions.}
\end{table}

However there are known limitations to the Common Task Framework - at least as described by Donoho. Data leakage plagues the contest literature as noted in \cite{Kaufman2011LeakageAvoidance} and \cite{Narayanan2011LinkChallenge}.\footnote{In the context of a CTF data leakage refers to ``cheating'', deliberate or accidental. For example training a model using data that would not in the real world be available to the model such as future values of an exogenous time series.} The CTF is primarily a research not a production paradigm, and cannot be expected to help practitioners materially with model deployment, ongoing model relevance or model oversight. 

For example, if a contest is run and a model chosen - what is the contest organiser to do when new data becomes available or a regime change occurs? For these reasons, we do not think the capabilities of talented individuals around the world are being fully leveraged. 

\section{Our experience crowd-sourcing statistical ``lambdas'' in a large firm}
\label{sec:formal}

Turning to our project, we move the discussion to streaming data problems - in contrast to the historical data setup of the CTF. Every major business includes instrumented processes in need of prediction and ours is no exception. Our actual examples include diverse tasks such as estimating bank branch activity an hour ahead for many branches, predicting trading volume for corporate bonds, and estimates of fifteen minute ahead electricity consumption in municipal buildings. There are thousands of possibilities. 

In our project we force participants to provide models that work on real-time streaming data. The reasons for our focus on streaming data are:
\begin{itemize}
    \item This is what business needs
    \item Over-fitting is reduced
    \item Data leakage is eliminated
    \item Data search becomes part of the contest
\end{itemize}
To expand on the first point, notice that in the traditional CTF the burden of model deployment, data gathering, data cleaning and sundry practical issues falls on the creator of the contest rather than the participants - so it might reasonably be asked who is doing most of the work. 

To the second and third points, we believe it is important to evaluate purely out of sample. The CTF is particularly awkward in the time series setting because no matter how training data is chosen, some combination of data leakage (future data in the training) or staleness is all but inevitable. 

But most importantly the classic CTF presents a Faustian bargain. Does one want clever people to find relevant exogenous data or not? 
This benefit of crowd-sourcing can ``ruin'' an historical contest, as noted, because contestants who find causaly related data will also find the future values of the same. We turn this limitation into a feature.  

A simple protocol for ``Repeated Online Analytical Response'' (ROAR) demands that a consumer of a stream of micro-predictions and a producer interact in the following manner. 
\begin{enumerate}
    \item Consumer of micro-prediction sends a ``question'' to producer, typically a request for a prediction of a quantity that will be revealed by the passage of time.  
    \item Producer responds ``immediately'' with an prediction (say within a few seconds).
    \item Later, the consumer sends the producer the revealed ground truth so that the producer of prediction might learn.
    \item Later, the consumer sends the producer compensation - typically based on {\em relative} accuracy.\footnote{The consumer can execute this protocol with multiple producers.} 
\end{enumerate}

One economical class of producer is a ``lambda'' \cite{Adzic2017ServerlessImpact} - a program that wakes for a short time, responds, and then avoids the use of compute resources until the next invocation.

The repetition of this loop is important. There is no contract between consumer and producer. Instead we rely on the number of questions of the same type being large, with both parties playing a repeated game. 

We created a ROAR platform for streaming prediction contests and deployed it inside a large firm. We highlighted a bond trading volume contest for the purpose of launching the platform in March 2019. As each data point or data vector was revealed, a request for a prediction of the next value was sent to competing bots. 

In principle any employee of the firm could enter any micro-service satisfying a required signature and data model. In practice, creation of bots was facilitated by model deployment software and scaffolding code freeing the community of participants from the minutiae of cloud micro-service deployment.\footnote{Domino Data Labs' cloud data science platform was used.} 

While this pattern undersells the notion of a prediction economy it was seen as a valuable first step in creating an active community, and a trial of a the key hypothesis: model deployment can be democratized. Employees were able to enter this contest by registering model URLs and authorization keys with the contest application. The cumulative performance of bots was reported on a leader board. 

Most bots were authored in Python. A broadcast email stated that the winner would received an all expenses paid trip to a major Machine Learning conference.    

The streaming setup addressed some of the downsides of historical data contests.  Over-fitting is arguably less of a problem because all assessment is made out of sample. Similarly, data leakage is all but eliminated. 

Compared to the classical Common Task Framework, we expected participants to have more difficulty. Submitting a tabular data set is one thing, but deploying, maintaining and improving a model that must learn on the fly was perceived as a significant adoption hurdle. In particular we expected some push-back from those more familiar with an offline batch calibration paradigm - one which was likely to leave them at a disadvantage as new data points streamed in.    

\begin{figure}
\includegraphics[width=0.5\textwidth]{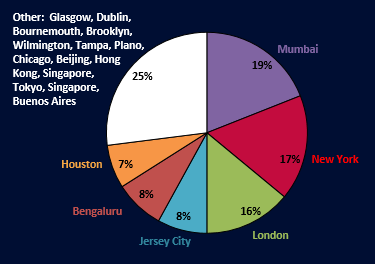}
\caption{Participation by city}
\label{fig:locale}
\end{figure}

Surprisingly the adoption was very good and well beyond our expectations. Participation over the course of several days very nearly overwhelmed the platform. Of the $1000$ employees who engaged in some way we found that $60$ percent interacted frequently and approximately $300$ completed all the required steps to participate in a meaningful manner. This number would have been much larger had we not closed entries. Neither geography nor timezone prevented bot creation, as shown in Figure \ref{fig:locale}. The vast majority of participants were successful without any human assistance.   

Interviews with participants revealed that contrary to expectation, participants enjoyed the additional difficulty. This included the technological challenges (such as storing and maintaining state) and statistical challenges (representing state, online learning and so forth). 

We observed spontaneous collaboration taking on various forms. The type that was most pleasing - given the larger ambition prefaced above - was bot stacking. Specifically one bot received the question and, before returning its answer, asked another bot to try to predict its error.\footnote{The secondary prediction of the error must, of course, occur fast enough to allow the receiving bot to reply to the original question.} 

Thus while we expected that this first experiment would be analogous to a ``mere'' perceptron in fact some depth was emerging even before we took steps to actively encourage it. The residual pattern was also applied with one bot calling itself. Clearly we underestimated the crowd's ingenuity. 

Another pleasant surprise was that stacking (chaining) of bots achieved technological goals as well as statistical ones. For instance a bot could call out to an in-house analytical system that some participants were comfortable with, or to a different language environment. Also within days, employees had created reference bots that satisfied the API from an alternative system without the need for this hop. 

Participants chipped in to help provide each other with common conveniences such as caching, storing and manipulation of historical questions. Some bots made calls to existing analytical endpoints. Automated Machine Learning was relatively easy to incorporate into bots despite the real-time nature of the contest. 

Participants collaborated on the creation of docker environments stocked full of popular analytical Python libraries, obviating the need to start from scratch and spend time pip installing everything from xgboost to auto-keras. References to relevant literature was shared between participants.  

Participants and onlookers came up with new uses of the ROAR protocol, for both commercial and philanthropic purposes. One of the discoveries of this exercise was that usage patterns could be better advertised. For instance the term prediction is loaded, and tends to obscure ROAR applications such as data cleaning, recommendation, anomaly detection and so forth.

The explicit nature of the ROAR protocol as stated above suggests that only supervised learning problems are appropriate, whereas in fact there is nothing in the protocol that prevents the ground truth message from using the responses. Schemes such as expectation maximization are readily adapted to this situation.  

Similarly the application of the ROAR protocol to reinforcement learning and optimal control problems may not be immediately apparent. Nothing in the protocol prevents action-conditional prediction and in addition, ROAR facilitates new approaches to this problem. For instance ROAR can achieve a separation of concerns (specialization) by having one contest stream for a value function and another an advantage function. For added flexibility the value function target can be defined with telescoping differences - as with the temporal difference learning trick in \cite{Sutton1988LearningDifferences}.\footnote{A catalog of ROAR patterns is forthcoming.} 

Another concern raised by stakeholders was privacy of data. At the time of writing this is being addressed in a rather direct way through the establishment of privacy preserving agents using a combination of techniques from one-time pads to multi-party computation. By proving that a large number of employees can actively use data for prediction without learning anything about the data, we hope to change dramatically the way in which confidentially and prediction are viewed.\footnote{Advanced statistical patterns can be employed such as the use of synthetic data for algorithm recruiting. But there are also simple techniques such as white-listing, or using ROAR to predict quantities that are causally related to, or ingredients to models for, the quantity of actual interest.}  

The contest provided a purely objective measure of performance absent the hierarchical management of talent that this article seeks to challenge. The winner was not drawn from those claiming expertise in the domain area nor employed within proximity to the trading desks. Instead, it was a relatively junior employee with only a years' experience working on an unrelated task from an office in Mumbai. 

One front office group responsible for predicting similar quantities - with over a decade's experience in the market - discussed electing a representative to challenge for the title - though this did not eventuate. Given the potential discomfort the event was generally received with good humour. 

Another finding was that the success of prediction algorithms seemed in the short term to be largely uncorrelated with algorithm ``buzz'' - though to formalize this statement might present some methodological challenge. The winner used a sensible stacking of workhorse models - logistic regression and ridge regression - with some custom feature engineering. 

That said, after the contest period was over and another week or so of data had flowed through a bot using a temporal convolutional network started to approach similar accuracy with less human tinkering. This type of progression was in line with expectations, and brings out a key benefit of the setup. Time to market is always top of mind for businesses and in the traditional workflow this can demand a \Qidone{trade-off} between hare and tortoise models. No so here. 

There were knock-on effects of running the contest which were not picked up by our metrics. Some employees were not ready to enter the contest due to lack of experience with Python programming or statistics - but the event and spontaneous community creation around it provided them with the impetus to begin learning. It will be very gratifying to see some of these employees participate in future events. A structured collection of education materials including references and videos was created to assist this path. 

Associated with this development, feedback from participants included many feature requests of different types with a common goal of increasing accessibility. For example, one participant requested the ability to use Excel to enter a contest - a suggestion we take seriously. Some very senior leaders participated but their time commitments prevented them from directly programming. However they were able to convey intent to proxy programmers based on their market knowledge. 

In the firm in which this experiment took place, there were technology hubs in cities around the world. One such hub was particularly active in ways that extended well beyond the creation of bots. They helped directly in the development of the platform. They organized biweekly educational sessions. They identified many new commercially significant micro-prediction streams that are amenable to the ROAR pattern of usage, and they went so far as to produce marketing collateral including videos which were shown to leadership. 

The reaction to the experiment was pleasing. Many employees expressed a strong belief in the broad direction and in the specific use as a connectivity tool between far flung hubs. After strengthening those bonds it seems likely that the paradigm can be lifted into communities near those hubs. Some employees identified applications in health-care at a local hospital. Others in transport. Coincident with the contest one forward looking city was already preparing to use ROAR in earnest and designating rack space in a town owned data center. 

\section{Future potential of the ROAR protocol}

Greatly encouraged by our empirical results we feel more confident in espousing the bigger picture, which was the original intent of creating a multi-agent system glued together with the ROAR protocol. Participants can do more than participate in a contest. They can be a contest - or something better.

It would be philosophically inconsistent to suggest that streaming contests are a great way to predict and yet not grant this ability to participants (as one of several tools they can use for this purpose). For example, if a participant requires precipitation as a model ingredient for predicting travel delays - why would they not consider using the ROAR protocol to predict precipitation? 

We adopt the terminology ``agent'' to refer to a bot that both answers questions and asks them. Agents, which provide streams of predictions of quantities of civic, economic or scientific relevance, can be viewed as nodes in a new kind of crowd-sourced Prediction Web - with the idea that this lattice of micro-nowcasts might one day grow to a size where it is indispensable for real-time decision making.

In any economic system, including our special variety, orchestration of the supply chain can be viewed as a variety of message passing. Changes to a single number, price, drive decisions made by the owner of a value adding firm. For example she may increase of decrease purchases or sales, find substitutes or search for new buyers. Propagation of the price of tin, Hayek's famous example, leads to adjustments in use across the globe. This feat of orchestration unthinkable by other means (such a cold calling users of tin). Hayek remarked that this elegance is often overlooked. \cite{Hayek1945TheHayek}
\begin{quote}
I am convinced that if it were the result of deliberate human design, and if the people guided by the price changes understood that their decisions have significance far beyond their immediate aim, this mechanism would have been acclaimed as one of the greatest triumphs of the human mind. 
\end{quote}
The caveat is economic frictions of trade, but it will be clear to the reader that we have designed this economy from the outset to be virtually frictionless.  

Today there are no supply chains for micro-predictions. The manner in which humans currently go about achieving micro-prediction is preindustrial in this sense, as with the labor of a master craftsman taking raw materials to a final product versus a production line worker achieving much greater throughput. To us this does not represent a compelling solution to the central problem of economics. Small wonder that small to medium size businesses, not for profits and individuals cannot afford the product. Not everyone can employ in-house data science artisans. 

Accessibility to commercial supply chains is the current blocker - not a shortage of talent. Anyone should be as free to participate as they are to edit Wikipedia. Someone will write an ingenious agent that works on many different micro-prediction tasks and does a better job of utilizing other agents. Certainly anyone should be able to establish a contest.  

When agents ask other agents sub-questions, the humble contest mechanism will not win out. For instance an agent should be able to make use of useful orthogonal information from another agent even if this prediction is not the most accurate. This is true even if the sub-question is precisely the same as the original question - which it need not be. We observed in our contest, for example, that participants well down the leader board were statistically significant in an ensemble model.   

For this reason the task faced in the design of an agent is more akin to the general problem of generalized regression than contest design. The additional complication is that the reliability of inputs is cost sensitive. As this problem encompasses most of statistics and Machine Learning, and represents an endless challenge, it is undoubtedly a task for the crowd. Who can design the best ``router'' for the Prediction Web?  

\section{An errors-in-variables perspective on agent and protocol design}

In our economy statistical methods are capital goods. They convert micro-prediction streams into more valuable ones. To illustrate some possible issues in agent and protocol design we assume in what follows that the agent employs an errors-in-variables model. \cite{Hall2009MeasurementPerspective},\cite{Buonaccorsi2010MeasurementApplications}. 

From this vantage, agents can be thought of as buyers and sellers of precision. A reasonable question to ask is whether communication of pricing should be part of the protocol. 

To formalize the discussion we presume $n$ streams of regularly sampled univariate data denoted $x^{(1)}$ through $x^{(n)}$. For each $i=1,\dots,n$ the sequence $x^{(i)}_1,x^{(i)}_2,\dots$ is called a ground truth.
Further we suppose that $\hat{x}^{(i)}_1,\hat{x}^{(i)}_2,\dots$ are the point estimates supplied by the $i$'th bot and that bot $i$ supplies an unbiased point estimate with independent normally distributed errors whose variance is $\sigma_i^2$. We denote the precision $p_i=1/\sigma_i^2$ and we shall think of precision as the unit of quality. For emphasis: $p_i=p_i(c_i)$ to indicate that the precision of the $i$'th bot will be more accurate over time if paid more. 

Bots $1$ through $n$ will be termed {\em children}. A {\em parent} is now introduced that consumes the predictions of the $x$'s created by the child bots and uses these to predict some other quantity $y$. It is this parent agent that will decide compensation $c=(c_1,\dots,c_n)$ paid to bots $1$ through $n$. The parent will in turn receive compensation depending on the precision of its own predictions of ground truths $y_1,y_2,\dots$. We denote these predictions, which are assumed to be contemporaneous with the child predictions, by $\hat{y}_1,\hat{y}_2,\dots$. Thematically we represent parent child relationship and predictions with errors as follows.

\begin{tikzpicture}[auto, outer sep=3pt, node distance=2cm,>=latex']
\node [] (y) {$\hat{y}_j$};
\node [, right of = y] (X) {$\hat{\beta}$};
\node [,above right of = X] (x1) {$\hat{x}^{(1)}_j=x^{(1)}_j + \eta^{(1)}_j \ \ \ \ \eta^{(1)}_j \sim N(0,\sigma_1^2)$ };
\node [,below right of = X] (x2) {$\hat{x}^{(2)}_j=x^{(2)}_j+\eta^{(2)}_j\ \ \ \ \eta^{(2)}_j \sim N(0,\sigma_2^2)$};
\draw [<->,thick] (y) --  node {}(X) ;
\draw [<->,thick] (X) --  node {}(x1) ;
\draw [<->,thick] (X) --  node {}(x2);
\end{tikzpicture}        \hspace{20mm}

These diagrams can grow to the right, since the $i$'th bot producing $\hat{x}_i$ might in turn be using the results of other prediction bots. They can grow to the left if $\hat{y}$ becomes an ingredient for some other prediction. Graphs might recombine and in generality, a parent bot might produce multiple outputs. The salient point here is that the agent performs a selfish local optimization just as a value adding firm in an economy pays for inputs and is paid for output. 

Ideally, agents are keenly aware of their shadow prices of precision and whether to pay their children more or less. Given knowledge gleaned over time about the best way to combine inputs, this seems like a routine calculation provided there is sufficient streaming data to allow high quality estimation of the model parameters. For instance we might assume millions of observations $x_j$ and $y_j$ per week, and assume the parent changes compensation schemes and coefficients over the weekend. 

It is difficult to predict the equilibrium or transient behaviour of agents. However we are brave enough to suggest that a simple heuristic such as affine precision pricing might help in agent design. The parent could base a strategy around the child's precision following the rule: 
\begin{equation}
\label{eqn:affine}
     p_i(c_i) =  const + \frac{1}{\rho_i} c_i 
\end{equation}
in response to payment $c_i$. We used $1/\rho_i$ for the coefficient so that $\rho_i$ plays the role of the child's price of precision. 

This argues for including price communication in the protocol (namely $\rho_i$) either from parent to child or vice versa. But would these prices be stable enough to be useful? 

A supply side plausibility argument for stability of $\rho_i$ might appeal to stylized production examples. For instance if the child arrives at her estimate by means of collecting independent unbiased data points with independent measurement errors around a ground truth, at a fixed cost per data point, then the child's manufacturing cost of precision is indeed linear. 

But does this mean the parent's pay scale will also exhibit roughly constant $\rho_i$? To us the tight coupling between parent and child which may evolve over the repeated ROAR game resembles non-market economic arrangements. In the multi-period principal agent model \cite{Holmstrom1987AggregationIncentives} for example it is demonstrated that a linear compensation scheme emerges as optimal in a multi-period setting. 

Again, this is a speculative supply side argument and of greater relevance might be actual applications where demand for precision is roughly linear. For example in a stylized model of market making the profitability of a market maker falls as $exp(-\sigma)$ where $\sigma$ is the standard error in the estimate of a key sufficient statistic.\footnote{Details upon request.} 

It would seem to benefit to child to know the parents bid for more precision - assuming this is not so wildly fluctuating as to be useless.  

Agents can be designed with straightforward precision calculations. Consider the matching pursuit algorithm \cite{Mallat1993MatchingDictionaries}. This can be implemented by an agent which performs one projection and sends residuals to itself - or to the crowd which includes itself. The variance in residuals contributes to the parent's prediction variance in a straightforward linear manner.  

These considerations suggest that precision pricing is so useful it should be codified in the protocol. On the other hand it can be relegated to a heuristic. To argue against the inclusion of precision price signaling in a formal protocol, which is otherwise uncomplicated, we can attack the generality of our errors-in-variables framing. There are many complexities to the repeated game.

For example, even for linear agents the statistical concept of attenuation complicates the picture. To illustrate assume that both the true relationship and parent model are both linear. 
\begin{equation}
\label{eqn:yj}
        y_j = b_0 + \sum_{i=1}^n b_i x^{(i)}_j
\end{equation}
The true coefficients are collectively labelled $b=(b_0,b_1,\dots,b_n)$. It is well known from the errors in variables literature that the parent's estimates $\hat{b}_i$ will be attenuated due to the noise in $\hat{x}_i$.\footnote{Attenuating the coefficients is necessary so that $\hat{y}$ is unbiased.}

Say within a long epoch the parent asymptotically learns the true coefficients $b$, irrespective of the child precision $\{p_i\}_{i=1}^n$.\footnote{Identification is a lurking issue, as discussed in \cite{Bekker2006CommentModel}.  
}
Introducing notation $\gamma$ we will have
$
  \hat{b}_i \rightarrow \gamma_i b_i\
$
where $\gamma_1,\gamma_2,\dots$ are shrinking parameters. This is because the parent uses the estimate
$$
        \hat{y}_j = \hat{b}_0 + \sum_{i=1}^n \hat{b}_i \hat{x}^{(i)}_j \rightarrow  \gamma_0 {b}_0 + \sum_{i=1}^n \gamma_i b_i \hat{x}^{(i)}_j
$$
with attenuated coefficients in order to avoid bias in $\hat{y}_j$. Subtracting this from equation \ref{eqn:yj} we see that there will be error in $\hat{y}_j$ not purely attributed to weighted sums of errors in the ingredients $x^{(i)}_j$. 

Thus parent strategy is non-trivial due to attenuation. There is incentive to pay to discover more about the true coefficients in the short term, for instance. 
\section{Summary}

While the general problem of prediction will never have a universal solution and is inherently hard and expensive, we have argued that quality micro-prediction should be inexpensive and ubiquitous. The rise of \Qidone{m}achine \Qidone{l}earning belies this distinction - notwithstanding the importance of new techniques \cite{Breiman2001StatisticalCultures}. Machine Learning advances have been catalyzed by the arrival of a large number of problems where there is sufficient data to discern good models from bad by means of a simple score. But if the assessment of the quality of a prediction is feasible by fully automated means, why not the management of the entire process? 

We invite those with expertise in automated machine learning to consider this algorithms-as-managers problem (noisy inputs that respond to compensation). Cost-aware generalized regression agents can play the role of Hayek's man-on-the-spot. They perform their selfish decision making and unwittingly improve a supply chain. 

In the spirit of this workshop we invite the reader to participate in the creation of a growing reusable lattice of micro-predictions, and discuss pros and cons of protocol details.

We believe adoption will be driven on the demand side by business sponsors in our organization, but also civic and scientific uses. Due to sharing of public predictions the marginal cost of the $n+1$st prediction should fall. 

By calling this setup a Prediction Web we acknowledge the inspiration of Tim Berners-Lee and the web itself \cite{Tronco2010AInternet}. A crowd of bots following a ROAR protocol or similar represents a crowd-sourcing of the causal structure between real-time forecasts of quantities of general interest. This is directly analogous to the crowd-sourcing of backlinks between documents. That structure was a necessary precursor to efficient web search. So to, we believe that a lightweight prediction economy is a prerequisite to solving micro-prediction.  

Technically this is made feasible by continuing advances in model deployment software which free invidual contributors from the need for a DevOps team. Our experiment establishes that there is no serious barrier to entry for data scientists and technologists looking to add value via the ROAR protocol. 

At minimum a collection of competing selfish prediction lambdas with disparate authors is an interesting type of ``Supermind'' \cite{Malone2018HowWork} and an attempt to tackle what Hayek considered the true challenge of economics:
\begin{quote}
    The problem is precisely how to extend the span of our utilization of resources beyond the span of the control of any one mind; and, therefore, how to dispense with the need of conscious control and how to provide inducements which will make the individuals do the desirable things without anyone having to tell them what to do.
\end{quote}

Our lambda economy anticipates the main objection to any free market argument - frictions of trade \cite{CoaseR1937NatureFirm}. The typical cost of negotiating an enterprise data feed, hiring a quant or other mechanics of the extant economy exceed the frictional costs of trade in our setup by many, many orders of magnitude.   

If we are right the economic forces operating on this microscopic level will come to become the dominant orchestrating principle in the supply chain of micro-prediction. And if so, it will be ironic that the job of managing AI, or at least a large class of applications, will be one of the first to be replaced by AI.

\section*{Acknowledgements}
This work was funded by the Corporate and Investment Banking division of JPMorgan Chase \& Co.\footnote{ Opinions expressed are those of the author alone. Opinions and estimates constitute our judgment as of the date of this Material, are for informational purposes only and are subject to change without notice. This Material is not the product of J.P. Morgan’s Research Department and therefore, has not been prepared in accordance with legal requirements to promote the independence of research, including but not limited to, the prohibition on the dealing ahead of the dissemination of investment research. This Material is not intended as research, a recommendation, advice, offer or solicitation for the purchase or sale of any financial product or service, or to be used in any way for evaluating the merits of participating in any transaction. It is not a research report and is not intended as such. Past performance is not indicative of future results. Please consult your own advisors regarding legal, tax, accounting or any other aspects including suitability implications for your particular circumstances. J.P. Morgan disclaims any responsibility or liability whatsoever for the quality, accuracy or completeness of the information herein, and for any reliance on, or use of this material in any way.} We are grateful to sponsors Samik Chandarana and Michael Grimaldi representing the Analytics and Technology organizations respectively. We thank direct and indirect members of the ROAR team including consulting firm Giant Machines and partners Domino Data Labs, organizers of a senior leaders conference where ROAR was promoted, internal communications, data science management and employees in our technology hubs - singling out Houston. We thank Antoine Toussaint for leading development of the prototype. We are grateful to the hundreds of employees who continue to participate in various ways as noted in the body. This document benefited from thoughtful feedback from referees and from Antigoni Polychroniadou in the JP Morgan Artificial Intelligence group. We acknowledge research funding for ongoing ROAR study at Stanford University from that same group. We also thank members of the MIT Center for Collective Intelligence for numerous extensive discussions.

\bibliography{ROAR_bib}
\bibliographystyle{icml2019}

\end{document}